\documentclass[trackchanges]{aastex7}

\usepackage{booktabs}
\usepackage{array}
\usepackage{threeparttable}
\usepackage{multirow}

\begin{document}

\title{Gamma-ray Emission from the S147 Region: Indication of Escaping Cosmic Rays Interacting with Molecular Clouds}

\author[sname=Yang]{Huan Yang}
\affiliation{Key Laboratory of Dark Matter and Space Astronomy, Purple Mountain Observatory, Chinese Academy of Sciences, Nanjing 210023, China}
\affiliation{School of Astronomy and Space Science, University of Science and Technology of China, Hefei 230026, China}
\email{huanyang@pmo.ac.cn} 

\author[orcid=0000-0002-5965-5576, sname=Liu]{Bing Liu}
\affiliation{Key Laboratory of Dark Matter and Space Astronomy, Purple Mountain Observatory, Chinese Academy of Sciences, Nanjing 210023, China}
\affiliation{School of Astronomy and Space Science, University of Science and Technology of China, Hefei 230026, China}
\email[show]{liubing@pmo.ac.cn} 

\author[orcid=0000-0001-8500-0541, sname=Zeng]{Houdun Zeng}
\affiliation{Key Laboratory of Dark Matter and Space Astronomy, Purple Mountain Observatory, Chinese Academy of Sciences, Nanjing 210023, China}
\affiliation{School of Astronomy and Space Science, University of Science and Technology of China, Hefei 230026, China}
\affiliation{Key Laboratory of Astroparticle Physics of Yunnan Province, Yunnan University, Kunming 650091, People's Republic of China}
\email{zhd@pmo.ac.cn}

\author[orcid=0000-0002-2750-3383, sname=Huang]{Xiaoyuan Huang}
\affiliation{Key Laboratory of Dark Matter and Space Astronomy, Purple Mountain Observatory, Chinese Academy of Sciences, Nanjing 210023, China}
\affiliation{School of Astronomy and Space Science, University of Science and Technology of China, Hefei 230026, China}
\email[show]{xyhuang@pmo.ac.cn}

\begin{abstract}
We present a detailed analysis of $\gamma$-ray emission from the middle-aged supernova remnant (SNR) S147 (G180.0$-$1.7) using approximately 16.5 years of \textit{Fermi}-LAT data. 
Spatially, a new extended $\gamma$-ray component distinct from the emission associated with the H$\alpha$ filaments of the SNR shell is
identified. This new component exhibits a strong spatial correlation with dense molecular clouds (MCs) identified in CO emission at Local Standard of Rest velocities of $0$--$5\,\mathrm{km\,s^{-1}}$. 
Spectrally, the cloud-associated emission implies an underlying cosmic-ray (CR) proton population described by a hard power-law with an index of $\Gamma \approx 2.1$, compatible with the standard diffusive shock acceleration prediction. We interpret the $\gamma$-ray emission in this region with a hadronic scenario involving two distinct CR populations: trapped CRs reaccelerated within the radiative SNR shell as proposed in previous work, and escaping CRs illuminating the nearby MCs. 
The derived CR proton intensity in the MC region significantly exceeds the local Galactic background measured by AMS-02, consistent with the interpretation that the cloud is illuminated by particles accelerated by S147. These findings provide observational support for a CR-escape scenario during the earlier evolutionary phases of this middle-aged SNR and highlight the S147 MC component as a potential candidate for detection at TeV energies by LHAASO.
\end{abstract}

\section{Introduction}
\label{sec:intro}

Cosmic rays (CRs) were discovered more than a century ago, yet their origin remains one of the most fundamental questions in high-energy astrophysics. The ``knee'' in the all-particle CR spectrum around $3\times10^{15}\,\mathrm{eV}$ marks a transition in acceleration mechanisms or source populations, below which CRs are widely believed to be of Galactic origin \citep{Hillas2005}. Among the proposed Galactic accelerators, supernova remnants (SNRs) have long been considered the primary candidates, capable of efficiently accelerating particles to high energies via the diffusive shock acceleration (DSA) mechanism \citep{Bell1978,Blandford1987,Blasi2013}. The observed $\gamma$-ray emission from the vicinity of SNRs generally originates from two fundamental processes involving these accelerated particles: leptonic and hadronic \citep{Aharonian2013,2019ApJ...874...50Z}. In the leptonic scenario, relativistic electrons produce $\gamma$-ray through inverse Compton (IC) scattering off ambient photon fields or via non-thermal bremsstrahlung when interacting with gas. This mechanism tends to dominate in SNRs expanding into low-density environments, as exemplified by RX~J1713.7$-$3946 and RCW~86 \citep{Aharonian2006, Ajello2016}. In contrast, collisions between accelerated protons and the interstellar medium produce neutral pions that subsequently decay into $\gamma$-rays ($\pi^{0}\rightarrow2\gamma$). This hadronic process yields a characteristic spectral feature known as the ``pion-decay bump'' in the $50$--$200~\mathrm{MeV}$ energy range \citep{Dermer1986,Ackermann2013} and is prominently observed in middle-aged SNRs interacting with dense molecular clouds (MCs), such as W44 and IC~443 \citep{Ackermann2013}. 
With the accumulation of observation data, more and more gamma-ray emission from MCs interacting with or nearby SNRs, e.g., Kes41 \citep{liub_kes41}, W51C \citep{Jogler2016}, Kes 73 \citep{liub_kes73}, Kes 78  \citep{Shen2025kes78}, G335.2+0.1 \citep{Huang2025G335} were discovered.

In the hadronic scenario, the spectral and spatial characteristics of the $\gamma$-ray emission depend critically on the confinement status of the interacting protons and the distribution of target material. Particles accelerated at the shock front can either remain \textit{trapped} within the SNR shell or \textit{escape} into the surrounding medium. The trapped population, confined by magnetic turbulence near the shock, typically exhibits a source spectrum reflecting the acceleration process. In contrast, the escaping population propagates into the surroundings via diffusion. Because the diffusion coefficient is energy-dependent ($D(E) \propto E^{\delta}$), high-energy particles escape more efficiently and traverse the distance to remote clouds faster than low-energy ones \citep{Aharonian1996,Ptuskin2005,Gabici2007}. Consequently, 
MCs located at a distance from the shock are illuminated primarily by these higher-energy escaping particles, leading to a $\gamma$-ray spectrum that is often harder than that of the shell-confined emission and the ambient Galactic CR sea, with the detailed spectral shape depending on the diffusion distance and time \citep{Gabici2009,Li2012,Nava2016}.

Simeis~147 (S147, G180.0$-$1.7) is a large, evolved shell-type SNR \citep{1969ApJ...155...67D} located toward the Galactic anticentre, with an angular diameter of approximately $180'$ \citep{Green:2019mta}. The remnant is renowned for its prominent filamentary morphology in H$\alpha$ line emission \citep{1976SvA....20...19L}, which correlates well with radio observations \citep{1980PASJ...32....1S, Xiao:2008mt}. Spatially associated with the remnant is the pulsar PSR~J0538$+$2817, located approximately $40'$ west of the geometrical center. Its parallax distance of $1.3^{+0.22}_{-0.16}$~kpc \citep{Chatterjee:2009ac} is consistent with the recent measurement of $1.37^{+0.10}_{-0.07}$~kpc derived from background stars \citep{Kochanek:2024lal}, establishing a reliable distance scale for the system. Consequently, we adopt a distance of $1.3$~kpc for all physical parameter estimations throughout this work.
Regarding the evolutionary state of S147, while standard Sedov-Taylor models suggest an older age \citep{Kochanek:2024lal}, recent X-ray observations by SRG/eROSITA support a supernova-in-a-cavity scenario with an age of approximately 35~kyr \citep{Khabibullin:2024hbr,Michailidis:2024lvb}. This revised age is consistent with the kinematic age of the associated pulsar \citep{Kramer_2003}, providing a coherent picture of the remnant's history. Crucially for CR propagation studies, 3D dust extinction mapping has identified a large dust structure at a distance of $d=1.22\pm0.21$~kpc, which spatially coincides with the SNR \citep{Chen2017dust}. Together with MCs identified in the Local Standard of Rest (LSR) velocity range of $-14$ to $+5\,\mathrm{km\,s^{-1}}$ \citep{Jeong2012}, these structures provide a potential reservoir of target material for CRs. While conclusive dynamical evidence for shock-cloud interaction (such as OH masers or line broadening) has been elusive, this complex environment is suitable for testing particle escape scenarios.

In the high-energy regime, previous studies using \textit{Fermi} Large Area Telescope (LAT) data detected a spatially extended $\gamma$-ray source ($0.2$--$10$~GeV) coincident with S147 \citep{Katsuta:2012zz}. The emission morphology in this energy band shows a possible spatial correlation with the prominent H$\alpha$ filaments, leading to the conclusion that the emission is dominated by CRs interacting with the dense gas within the filaments. However, observations above 10~GeV reveal a different picture. In this higher energy band, a simple flat disk template provides a better fit to the data than the H$\alpha$-based template used by \citet{Katsuta:2012zz}. Furthermore, this high-energy component exhibits a hard spectrum with an index of approximately 2 \citep{Fermi-LAT:2017sxy,Fermi-LAT:2017mzh}, which contrasts significantly with the softer spectrum measured at lower energies \citep{Katsuta:2012zz}. These discrepancies in both morphology and spectral index strongly suggest the presence of a distinct emission component in the region, separate from the CRs trapped in the filaments. Given the complex environment surrounding S147, where MCs have been detected in the vicinity \citep{Jeong2012} and dust structures suggest dense material at the remnant's distance \citep{Chen2017dust}, this hard component may arise from escaping particles accelerated by the remnant illuminating nearby molecular material. The diffusion of these escaping CRs into the surrounding clouds would generate significant hadronic $\gamma$-ray emission with precisely such a distinct hard spectrum, warranting a dedicated investigation into unassociated residuals near the SNR boundary.

In this work, we present a detailed analysis of the $\gamma$-ray emission from S147 using approximately 16.5 years of \textit{Fermi}-LAT data. We combine these data with CO emission templates to systematically search for gas targets of escaping CRs. We identify a $\gamma$-ray component spatially correlated with MCs in the $0$--$5\,\mathrm{km\,s^{-1}}$ LSR velocity range, distinct from the H$\alpha$-correlated emission. We demonstrate that the CO template provides a statistically superior description of this component compared to geometric models, and that this result is robust against different background modeling assumptions. Finally, we interpret the $\gamma$-ray spectra using a unified hadronic scenario in which the H$\alpha$ component is associated with trapped CRs interacting with the SNR shell, while the CO component is associated with escaping CRs illuminating nearby MCs.

\section{Molecular clouds distribution around S147}
\label{sec:MCs}

Recently, a 3D dust extinction study by \citet{Chen2017dust} identified a prominent dust structure (the ``S147 dust cloud'') at a distance range of $\sim1.0$--$1.6$\,kpc, which is consistent with the SNR's distance. 
Several dense clumps within this dust structure exhibit a strong spatial correspondence with the CO features observed in the region. Assuming these MCs are associated with the dust structure and thus located in the immediate environment of S147, they could constitute ideal targets. Protons escaping the acceleration site could diffuse into these clouds, producing hadronic $\gamma$-ray emission detectable by \textit{Fermi}-LAT.

\begin{figure}[htbp]
\centering
\includegraphics[width=0.95\textwidth]{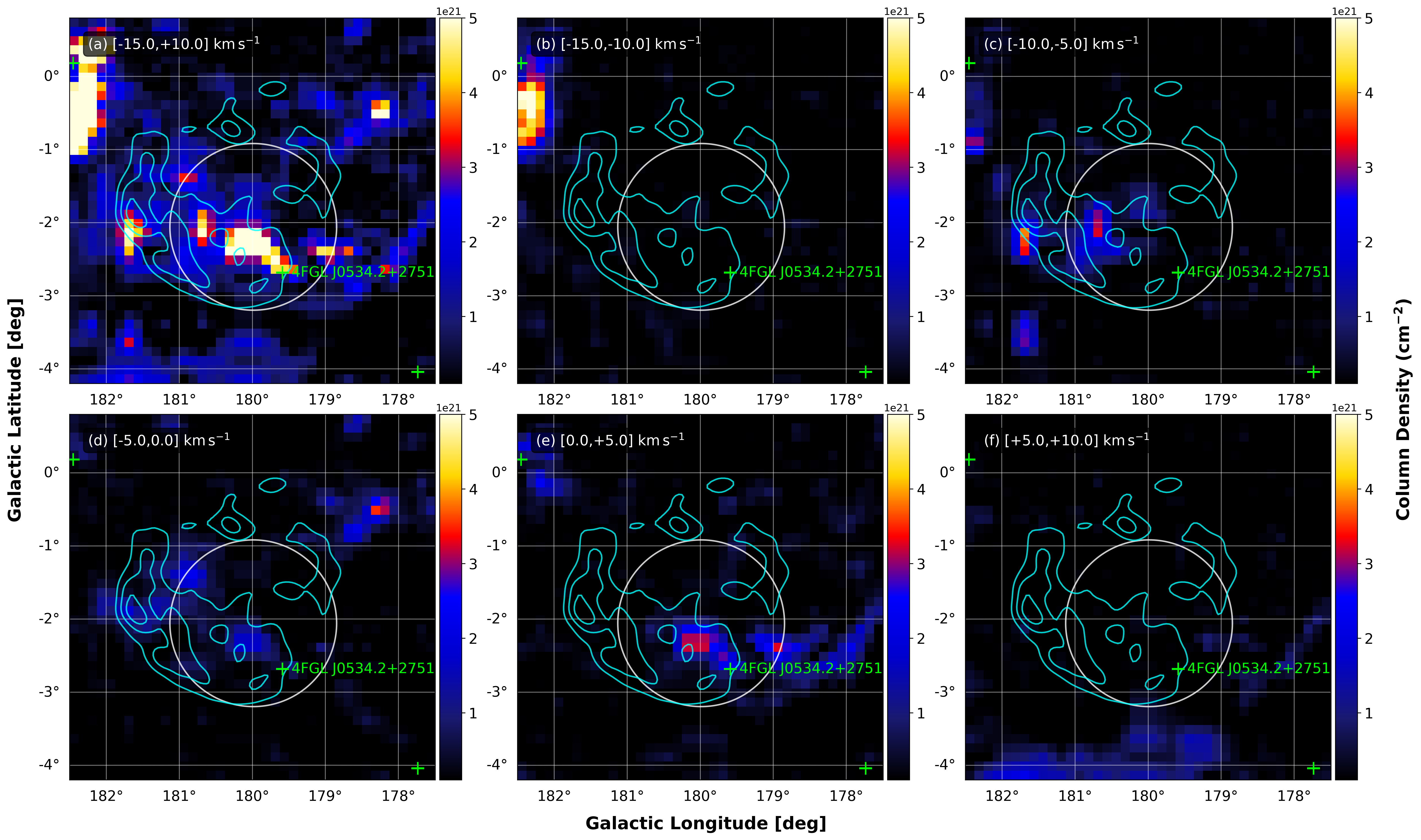}
\caption{Panels (a)--(f) show the hydrogen column density distribution in different LSR velocity intervals. Panel (a) covers the full velocity range from $-15$ to $+10~\mathrm{km\,s^{-1}}$, while panels (b)--(f) present consecutive $5~\mathrm{km\,s^{-1}}$-wide velocity bins within the same range. White circles denote the disk template adopted in the Third Catalog of Hard Fermi-LAT Sources (3FHL; \citealt{Fermi-LAT:2017sxy,Fermi-LAT:2017mzh}). Cyan contours outline the H$\alpha$ template proposed by \citet{Katsuta:2012zz}. Green crosses indicate the $\gamma$-ray sources listed in the 4FGL catalog.}
\label{fig:co}
\end{figure}

Given the location toward the Galactic anticenter, the kinematic distances of the MCs around S147 derived from Galactic rotation are highly uncertain \citep{Brand1993}. Thus, we prioritized morphological correspondence to identify the gas associated with the $\gamma$-ray emission. Using CO survey data from \citet{Dame2001}, we inspected the CO gas distribution across a broad velocity range of $v_{\rm LSR} \in [-15, +10]\,\mathrm{km\,s^{-1}}$, a range in which \citet{Jeong2012} found MCs that partially overlap with S147 along the line of sight.
Applying a conversion factor of $X_{\rm CO} = 1.8\times10^{20}\,\mathrm{cm^{-2}\,(K\,km\,s^{-1})^{-1}}$ \citep{Dame2001}, we derived the molecular hydrogen column density in different velocity intervals. The molecular gas distributions, overlaid with the contours of the H$\alpha$ emission from S147 \citep{Katsuta:2012zz}, are illustrated in Figure~\ref{fig:co}.
A comparison between the molecular gas morphology and the \textit{Fermi}-LAT source positions reveals a notable spatial correlation in the $0$--$5\,\mathrm{km\,s^{-1}}$ velocity interval (see Figure~\ref{fig:co}e). Specifically, the dense region of the MC in this range spatially overlaps with the unassociated source 4FGL~J0534.2+2751 \citep{4fgl-dr4}. Moreover, this molecular feature also coincides with the extended source 3FHL~J0537.6+2751e detected above 10~GeV \citep{Fermi-LAT:2017sxy}, which represents a hard-spectrum component with an extended disk morphology reported in previous analyses \citep[e.g.,][]{Fermi-LAT:2017mzh} and is indicated by the white circles in Figure~\ref{fig:co}. This spatial coincidence between the dense CO clump, the persistent unassociated source, and the reported high-energy extended emission may imply a physical association between the molecular gas and escaping CRs. 
Thus, motivated by this morphological correspondence, the molecular gas distribution in the $0$--$5\,\mathrm{km\,s^{-1}}$  velocity interval as presented in Figure~\ref{fig:co}(e) is adopted as the spatial template in the subsequent likelihood analysis. We further compare CO templates constructed from the full and sub-divided velocity intervals shown in Figure~\ref{fig:co} in Section~3.2, in order to assess whether this visually motivated template is also favored by the gamma-ray data.

\section{Fermi-LAT data analysis} 
\label{sec:analysis}

\subsection{Data Selection and Preparation}
\label{sec:data_prep}

In this work, we utilize approximately 16.5 years of \textit{Fermi}-LAT Pass~8 data, covering the observation period from August 4, 2008 to March 18, 2025. We select \texttt{SOURCE} class events (\texttt{evclass=128}) with both \texttt{FRONT} and \texttt{BACK} conversion types (\texttt{evtype=3}) in the energy range $100~\mathrm{MeV}$--$1~\mathrm{TeV}$. To minimize contamination from the Earth's limb, we apply a maximum zenith angle cut of $90^\circ$. Standard data quality selections are imposed using \texttt{(DATA\_QUAL>0 \&\& LAT\_CONFIG==1)}. Given the angular extent of S147, angular diameter of $\sim 3^\circ$, a sufficiently large analysis region is required to encompass the remnant and account for the Point Spread Function (PSF) containment. We therefore adopt a $14^\circ\times14^\circ$ square region of interest (ROI) centered on the geometrical center of S147 (J2000: R.A.\,$=84.75^\circ$, Decl.\,$=27.83^\circ$). We bin the data using a spatial pixel size of 0.1$^\circ$ and 10 logarithmic energy bins per decade. The baseline background model is constructed based on the \textit{Fermi}-LAT 14-year Source Catalog (4FGL-DR4; \citealt{4fgl-dr4}), including all point-like and extended sources located within $20^\circ$ of the ROI center. The Galactic diffuse emission is modeled using the standard template (\texttt{gll\_iem\_v07.fits}), and the isotropic diffuse background is described by the spectrum file (\texttt{iso\_P8R3\_SOURCE\_V3\_v1.txt}) provided by the \textit{Fermi}-LAT Collaboration. Energy dispersion effects are accounted for by enabling the correction on the spectrum (\texttt{edisp\_bins = -2}).

We perform binned maximum likelihood analyses over the selected energy range using the \texttt{Fermitools} package. Unless otherwise specified, the following fitting strategy is applied to all global likelihood analyses presented in this work (including Sections \ref{sec:morphology} and \ref{sec:spectrum}): the spectral parameters of sources located within $7^\circ$ of the ROI center with a detection significance $> 15\sigma$ are left free to vary; additionally, the normalizations of both the Galactic and isotropic diffuse emission components are treated as free parameters. All other spectral parameters are fixed to their catalog values to ensure fit stability. We verified that modest variations in these selection criteria (e.g., changing the significance threshold) do not significantly impact the separation of the H$\alpha$ and CO components or the spectral results.

\subsection{Morphological Analysis}
\label{sec:morphology}

To investigate the spatial distribution of the $\gamma$-ray emission, we focused on the energy range of $1~\mathrm{GeV}$--$1~\mathrm{TeV}$ for narrower PSF. We adopted the standard model from the \textit{Fermi}-LAT 14-year Source Catalog (4FGL-DR4) as our Baseline Model. In this catalog, the extended emission from S147 is modeled using an H$\alpha$ intensity map \citep{Katsuta:2012zz}, and the region includes the unassociated point source, 4FGL~J0534.2+2751, located near the western boundary of the remnant.

\begin{figure}[htbp]
\centering
\includegraphics[width=\textwidth]{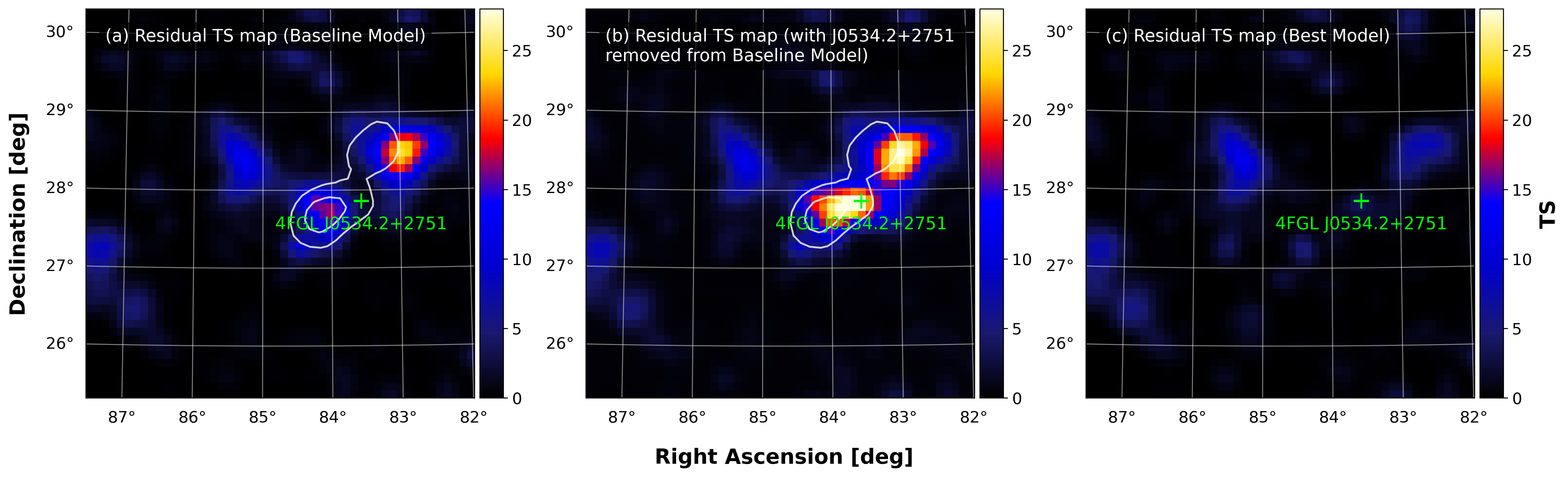}
\caption{
Residual TS maps of the $5^\circ \times 5^\circ$ region in the $1~\mathrm{GeV}$--$1~\mathrm{TeV}$ energy band. Green crosses indicate the $\gamma$-ray sources in the 4FGL-DR4 catalog, and white contours trace the MC distribution derived from the CO($0$--$5~\mathrm{km\,s^{-1}}$). Panel (a) shows the residual TS map obtained with the baseline model. Panel (b) shows the residual TS map calculated after removing the unassociated source 4FGL~J0534.2+2751 from the baseline Model, revealing the extended excess. Panel (c) shows the residual TS map for the best-fit model that includes the CO component while excluding 4FGL~J0534.2+2751. See Section\ref{sec:morphology} for more details.}
\label{fig:combined}
\end{figure}

We first examined the residual Test Statistic (TS) map of a $5^\circ \times 5^\circ$ region derived from this baseline model (\autoref{fig:combined}a). While the H$\alpha$ template generally describes the SNR filamentary emission well, we observed non-negligible residual excesses surrounding the unassociated source 4FGL~J0534.2+2751. The presence of significant residuals in the immediate vicinity of this unassociated source suggests that the simple point-source assumption may be insufficient. Instead, the point source and the surrounding excesses likely constitute part of a single, spatially extended structure which is not modeled properly. To visualize the intrinsic morphology of this extended feature without the bias of the point-source assumption, we removed 4FGL~J0534.2+2751 from the baseline model and calculated the residual TS map. The resulting map, shown in \autoref{fig:combined}b, reveals an extended $\gamma$-ray emitting structure that exhibits a notable spatial correlation with the dense clumps of MCs identified in the $0$--$5\,\mathrm{km\,s^{-1}}$ velocity range (overlaid as white contours in \autoref{fig:combined}b).
Motivated by this morphological correspondence, we replace 4FGL~J0534.2+2751 with an extended gamma-ray source (hereafter referred to as the CO component), modeling its surface brightness as proportional to the column density of the molecular gases in the $0$--$5\,\mathrm{km\,s^{-1}}$ velocity range and its spectrum as a simple power-law function.   
We then performed a maximum likelihood fit where the normalization and spectral index of the CO component were left free, and generated the residual TS map next based on the fitting results.
As shown in \autoref{fig:combined}c, the significant excesses observed previously are effectively suppressed. This indicates that the CO template provides a better description of the extended spatial morphology. 

To quantitatively validate this improvement, we performed a systematic comparison using the Akaike Information Criterion (AIC) \citep{1974ITAC...19..716A}. We define the improvement in AIC relative to the Baseline as:
\begin{equation}
\Delta \mathrm{AIC} = 2\Delta\ln\mathcal{L} - 2\Delta k,
\end{equation}
where $\Delta\ln\mathcal{L}$ is the increase in log-likelihood and $\Delta k$ is the change in the number of free parameters relative to the baseline model. 
Under this convention, a larger positive $\Delta \mathrm{AIC}$ suggests a preferred model description.

We tested two categories of spatial models to describe the excess emission. The first category is geometric models, aimed at testing whether the excess can be attributed to unresolved sources or to simple extended morphologies. To test the unresolved-source interpretation, we fitted models with one and two additional point sources while retaining 4FGL~J0534.2+2751. For the two-source case, we repeated the fit with the positions and power-law spectral parameters of both added point sources left free, using the results from the sequential procedure as initial values. The fit converged to the same best-fit values. We also tested a uniform disk template to examine the connection with the hard-spectrum emission detected at higher energies, with its position and radius fixed to those of the extended source 3FHL~J0537.6+2751e in the 3FHL catalog ($>10$~GeV), and a two-dimensional Gaussian model with free centroid and extension. The second category is a set of physical CO templates constructed over the velocity range $-15$ to $+10\,\mathrm{km\,s^{-1}}$ with intervals of $5\,\mathrm{km\,s^{-1}}$, to explore a possible association between the gamma-ray emission and molecular clouds along the line of sight.

Consistent with the visual inspection above, in all likelihood analyses the spectrum of the added spatial component was modeled as a simple power-law with free normalization and spectral index. In the spatial-model comparison, Scenario A denotes models in which 4FGL~J0534.2+2751 is retained and the listed component is added. This includes the tests with one and two additional point sources. Scenario B denotes models in which 4FGL~J0534.2+2751 is removed and the listed extended component is added, to test whether the new component can substitute for the cataloged point source.

\begin{table*}[htbp]
\centering
\caption{Comparison of Spatial Models for the S147 Region}
\label{tab:model_comparison}
\renewcommand{\arraystretch}{1.3}
\setlength{\tabcolsep}{6pt}

\begin{tabular}{l c cc cc}
\hline
\hline
\multirow{2}{*}{Spatial Model Added} & Spatial Parameters & \multicolumn{2}{c}{Scenario A: with J0534} & \multicolumn{2}{c}{Scenario B: w/o J0534} \\
\cmidrule(lr){3-4} \cmidrule(lr){5-6}
 & ($\alpha_{2000}, \delta_{2000}, r_{68}$) & $\Delta k$ & $\Delta$AIC & $\Delta k$ & $\Delta$AIC \\
\hline

None (Baseline) & - & $0$ & $0.0$ & $-4$ & $-45.6$ \\

\hline
\multicolumn{6}{l}{\textit{Geometric Templates}} \\

One additional point source & $82.95^\circ, 28.50^\circ, -$ & $+4$ & $22.7$ & $-$ & $-$ \\

Two additional point sources & $83.94^\circ, 27.77^\circ, -$ & $+8$ & $42.7$ & $-$ & $-$ \\

Disk (3FHL)$^{(a)}$ & $84.41^\circ, 27.86^\circ, 1.14^\circ$ & $+2$ & $42.3$ & $-2$ & $22.6$ \\

Gaussian$^{(b)}$ & $84.09^\circ, 27.77^\circ, 0.76^\circ$ & $+5$ & $53.2$ & $+1$ & $44.8$ \\

\hline
\multicolumn{6}{l}{\textit{Physical Templates}} \\

CO ($-15$--$+10\,\mathrm{km\,s^{-1}}$) & Fixed Map & $+2$ & $53.9$ & $-2$ & $41.3$ \\

CO ($-15$--$-10\,\mathrm{km\,s^{-1}}$) & Fixed Map & $+2$ & $-2.4$ & $-2$ & $-40$ \\

CO ($-10$--$-5\,\mathrm{km\,s^{-1}}$) & Fixed Map & $+2$ & $1.0$ & $-2$ & $-38$ \\

CO ($-5$--$+0\,\mathrm{km\,s^{-1}}$) & Fixed Map & $+2$ & $9.7$ & $-2$ & $-26.3$ \\

CO ($+0$--$+5\,\mathrm{km\,s^{-1}}$) & Fixed Map & $+2$ & $61.3$ & $-2$ & $\mathbf{66.6}$ \\

CO ($+5$--$+10\,\mathrm{km\,s^{-1}}$) & Fixed Map & $+2$ & $0.6$ & $-2$ & $-33.3$ \\

\hline
\hline
\end{tabular}

\medskip
\parbox{0.9\textwidth}{\footnotesize
\textbf{Note.}
Comparison of the goodness-of-fit for different spatial models.
\textbf{Scenario A} adds the listed component to the Baseline. \textbf{Scenario B} adds the listed component and removes 4FGL~J0534.2+2751.
$\Delta \mathrm{AIC} = \mathrm{AIC}_{\text{Baseline}} - \mathrm{AIC}_{\text{Model}}$. For the point-source tests, the listed point sources are additional sources added in Scenario A, where 4FGL J0534.2+2751 is retained.
The spatial extent is quantified by $r_{68}$, the radius containing 68\% of the source intensity.
(a) The Disk template is fixed to the location and size of 3FHL~J0537.6+2751e. Note that for a uniform disk of radius $R$, $r_{68} \approx 0.82 R$.
(b) The Gaussian template parameters are free in the fit. The value listed corresponds to the best-fit $\sigma = 0.50^\circ$, converted using $r_{68} \approx 1.51\sigma$.
The bold value indicates the best-fit model.
}
\end{table*}

The quantitative comparison (Table~\ref{tab:model_comparison}) yields three key results. First, simply removing 4FGL~J0534.2+2751 from the Baseline drastically degrades the fit ($\Delta \mathrm{AIC} = -45.6$), confirming the significance of the excess emission. Second, the geometric models improve the fit to varying degrees. In Scenario A, the one- and two-additional-point-source models improve the fit with $\Delta \mathrm{AIC}=22.7$ and $42.7$, respectively, but both remain less favored than the CO-based model. The Disk and Gaussian templates also improve the fit; however, their $\Delta \mathrm{AIC}$ values in Scenario B are lower than those in Scenario A (e.g., $44.8$ vs. $53.2$ for the Gaussian), suggesting that these simple extended morphologies cannot fully substitute for the cataloged point source. Third, the physical CO ($0$--$5\,\mathrm{km\,s^{-1}}$) template provides the best overall description. Among the tested models, it achieves the largest improvement, with $\Delta \mathrm{AIC}=66.6$ in Scenario B. The fact that this value is higher than that obtained in Scenario A ($61.3$) suggests that 4FGL~J0534.2+2751 may reflect a local enhancement within a more extended cloud-correlated emission structure, rather than a distinct isolated point source. Consequently, we identify the emission previously attributed to the unassociated source as part of extended hadronic emission from the cloud, and we adopt the model Baseline + CO($0$--$5\,\mathrm{km\,s^{-1}}$) $-$ 4FGL~J0534.2+2751 as the best-fit model for the subsequent spectral analysis. In addition, a visual comparison of the counts, model, and residual maps also supports this conclusion (see Appendix~\ref{app:counts_maps}).

\subsection{Spectral Analysis}
\label{sec:spectrum}

Adopting the best-fit spatial model from Section~\ref{sec:morphology}, we proceeded to characterize the spectral properties of the $\gamma$-ray emission around S147. 
We performed a global maximum likelihood fit over the full energy range of $100~\mathrm{MeV}$--$1~\mathrm{TeV}$ using the same fitting strategy described in Section~\ref{sec:data_prep}. In this configuration, the spectral shape of the extended SNR S147 (traced by H$\alpha$) was modeled as a LogParabola function, consistent with the 4FGL catalog definition, and the CO component was modeled as a simple power-law. The global fit yields a spectral index of $\Gamma_{\rm CO} = 2.1 \pm 0.05$ for the CO component, consistent with the hard component listed in the 3FHL catalog. Both components are detected with high significance: $\mathrm{TS}_{\rm H\alpha} \approx 274$ and $\mathrm{TS}_{\rm CO} \approx 147$, respectively.

To derive the spectral energy distributions (SEDs), we divided the energy range into 10 logarithmically spaced bins to perform bin-by-bin analyses.
The analysis at the lowest energies requires special treatment due to the large PSF of \textit{Fermi}-LAT, which causes severe source confusion between the spatially overlapping H$\alpha$ and CO components. To mitigate this, we merged the first two low-energy bins ($100$--$631$\,MeV) into a single interval and restricted the likelihood analysis to \texttt{evtype=32} (PSF3) events, utilizing the data class with the best angular resolution. This stringent selection minimizes the leakage of the SNR shell emission into the MC region caused by the broad PSF at low energies. Accordingly, the isotropic diffuse emission for this bin was modeled using the specific PSF3 tabulated spectrum (\texttt{iso\_P8R3\_SOURCE\_V3\_PSF3\_v1.txt}). For the subsequent higher energy bins, the standard data selection was applied. In each spectral bin, we performed a fit where the normalizations of both the H$\alpha$ and CO components, as well as the Galactic and isotropic diffuse backgrounds, were left free to vary. The spectral shape parameters were fixed to their global best-fit values to ensure stability within narrow energy bands. We imposed a detection threshold of $\mathrm{TS} \ge 4$ ($\sim2\sigma$). For energy bins where a component was not significantly detected, we calculated the 95\% confidence level upper limit.

\begin{figure}[htbp]
\centering
\includegraphics[width=0.6\textwidth]{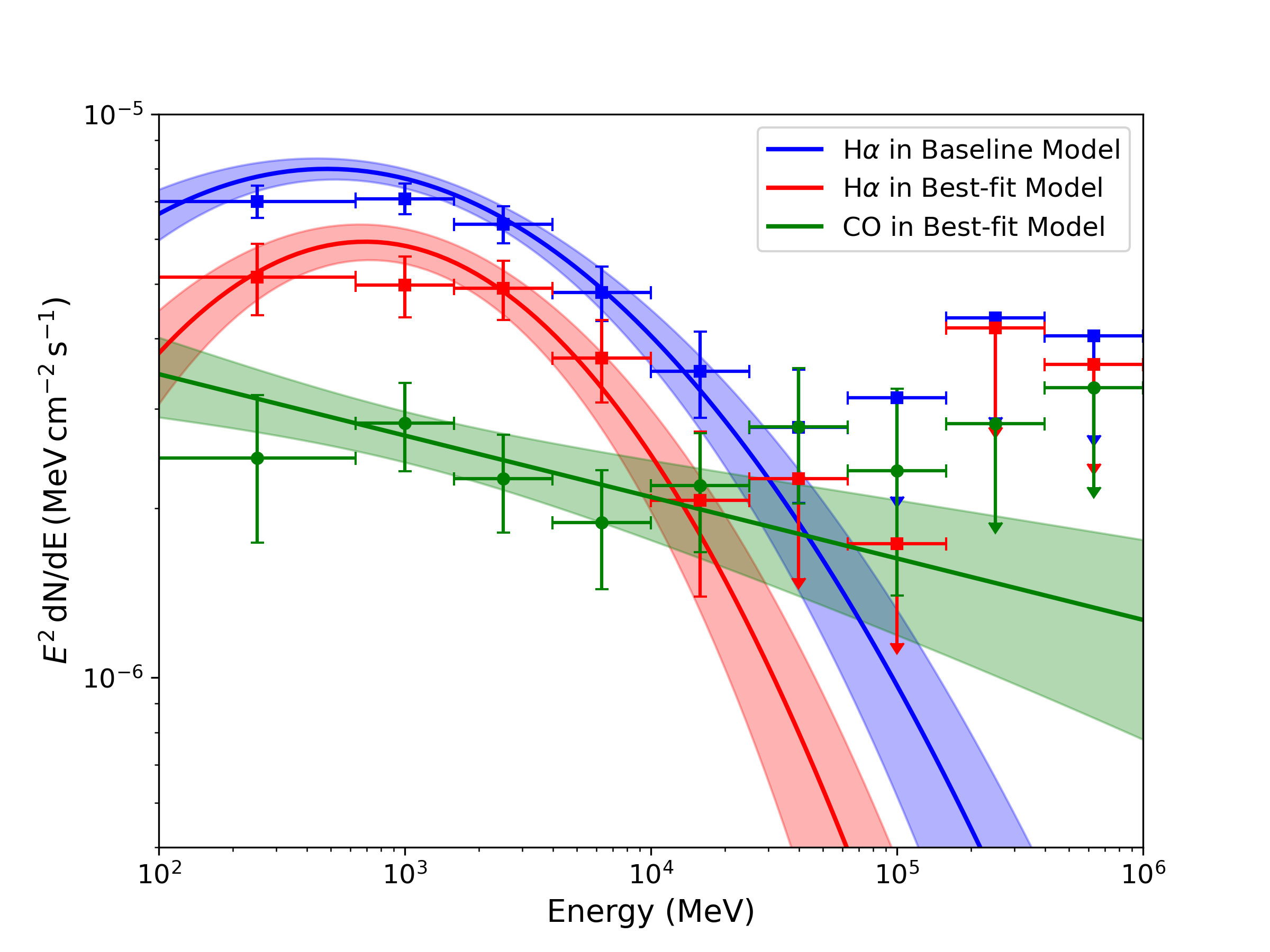}
\caption{SEDs of the $\gamma$-ray emission in the S147 region. 
The blue data points and shaded band represent the SNR H$\alpha$ emission derived from the Baseline model (without adding the CO component). 
The red and green data points represent the SNR H$\alpha$ and CO($0$--$5\,\mathrm{km\,s^{-1}}$) components, respectively, derived from the simultaneous fit in the optimal model. 
The solid lines and shaded bands indicate the global best-fit spectral functions (LogParabola for H$\alpha$, Power Law for CO) and their $1\sigma$ uncertainties.
Vertical error bars represent $1\sigma$ statistical uncertainties, and arrows indicate 95\% confidence level upper limits.
Note that the first data point represents the merged low-energy interval analyzed using exclusively PSF3 events.
}
\label{fig:SED}
\end{figure}

The resulting SEDs are presented in \autoref{fig:SED}. For comparison, we also derived the SED of the H$\alpha$ emission using the original baseline model (ie., excluding the CO component), shown as the blue curve. As evident in the figure, the inclusion of the CO component significantly reduces the flux attributed to the H$\alpha$ component (red curve) compared to the H$\alpha$-only case, suggesting that a significant portion of the emission previously modeled as SNR shell emission may be actually associated with the MCs. Furthermore, the CO component (green curve) exhibits a notably harder PowerLaw spectrum compared to the curved LogParabola spectrum of the SNR emission. This spectral distinction may support a scenario where energy-dependent CR escaping from the SNR illuminates the surrounding dense molecular material. The physical implications of these spectral features and the derived CR densities will be discussed in detail in Section~\ref{sec:discussion}.

\section{Discussion}
\label{sec:discussion}

The hadronic origin of the $\gamma$-ray emission in S147 has been previously established for the SNR shell component. This conclusion was supported not only by the striking spatial correspondence between the $\gamma$-rays and the H$\alpha$ filaments but also by the spectral consistency with the decay of neutral pions produced in proton-proton interactions \citep{Katsuta:2012zz}. In this work, the strong spatial correlation identified between the newly detected extended emission and MCs (Section~\ref{sec:morphology}) further reinforces the hadronic scenario, suggesting that $\gamma$-ray emission throughout the entire region, both the shell and the cloud interaction sites, results from inelastic collisions between CR protons and ambient gas. Building on this scenario, we proceed to derive the properties of the parent CR proton populations responsible for these emissions. Given the complex morphology of S147 and the uncertainties in the 3D separation between the accelerator and the clouds, we adopt a direct phenomenological approach that we constrain the CR proton density and spectral shape for both the SNR shell and the surrounding MCs using the SEDs derived in Section~\ref{sec:spectrum} in conjunction with the estimated total mass of the target gas.

The parent CR proton population is characterized by its differential intensity $J_{\mathrm{CR}}(E_{\mathrm{p}})$, which relates to the CR density via $\rho_{\mathrm{CR}} = (4\pi/c) J_{\mathrm{CR}}$. We assume the intensity follows a power-law form:
\begin{equation}
J_{\mathrm{CR}}(E_{\mathrm{p}}) = J_{0} \left( \frac{E_{\mathrm{p}}}{1~\mathrm{GeV}} \right)^{-\alpha},
\end{equation}
where $E_{\mathrm{p}}$ is the proton kinetic energy and $J_{0}$ is the normalization intensity. To relate the observed $\gamma$-ray emission to this parent population, we adopt the ``CR barometer'' formalism \citep[Eq.~1 in][]{Aharonian:2018rob}. In this framework, assuming CRs freely penetrate the clouds and are uniformly distributed, the expected $\gamma$-ray differential flux $F_{\gamma}(E_{\gamma})$ observed at Earth is given by:
\begin{equation}
F_{\gamma}(E_{\gamma}) = \frac{M_{\mathrm{gas}}}{d^2} \frac{\xi_{\mathrm{N}}}{m_{\mathrm{p}}} \int_{E_{\mathrm{p}}^{\mathrm{th}}}^{\infty} \frac{\mathrm{d}\sigma_{pp\to\gamma}}{\mathrm{d}E_{\gamma}}(E_{\mathrm{p}}, E_{\gamma}) J_{\mathrm{CR}}(E_{\mathrm{p}}) \, \mathrm{d}E_{\mathrm{p}},
\end{equation}
where $M_{\mathrm{gas}}$ is the total mass of the target gas, $d$ is the distance to the source, $m_{\mathrm{p}}$ is the proton mass, $\xi_{\mathrm{N}} \approx 1.8$ is the nuclear enhancement factor \citep{PhysRevD.90.123014}, $E_{\mathrm{p}}^{\mathrm{th}}$ is the kinematic threshold kinetic energy, and $\mathrm{d}\sigma/\mathrm{d}E_{\gamma}$ represents the differential cross-section for $\gamma$-ray production. This formulation depends solely on the ambient CR intensity $J_{\mathrm{CR}}$ and the mass-to-distance ratio $M/d^2$, eliminating the need for geometric volume assumptions. We acknowledge that the assumption of free penetration might be violated, as indicated by recent observational studies suggesting CR exclusion in dense MCs \citep{Huang:2020ngv,Yang:2023vza}. Such exclusion effects could effectively harden the observed $\gamma$-ray spectrum \citep{Gabici2007,Chernyshov:2024axd}. However, detailed modeling of such propagation effects is beyond the scope of this phenomenological study.

For the MC component, we adopt a distance of $1.3~\mathrm{kpc}$, consistent with that of S147. We estimate the gas mass from molecular hydrogen column densities derived from CO observations, focusing on the two dominant dense cloud structures with angular radii of $0.78^{\circ}$ and $0.50^{\circ}$. This yields a total molecular gas mass of approximately $8.5\times10^{3}~M_{\odot}$, which is treated as a fixed input to constrain the CR intensity. Regarding the SNR shell component, given the significant uncertainties in total mass estimates \citep{Katsuta:2012zz, Khabibullin:2024hbr}, we adopt a reference mass of $M_{\mathrm{shell, ref}} = 1000~M_{\odot}$. Since the normalization of the derived proton intensity scales inversely with the target mass ($J_{0} \propto M^{-1}$), the resulting CR spectrum for the shell is reported relative to this reference baseline.

\begin{figure}[htbp]
\centering
\includegraphics[width=0.7\textwidth]{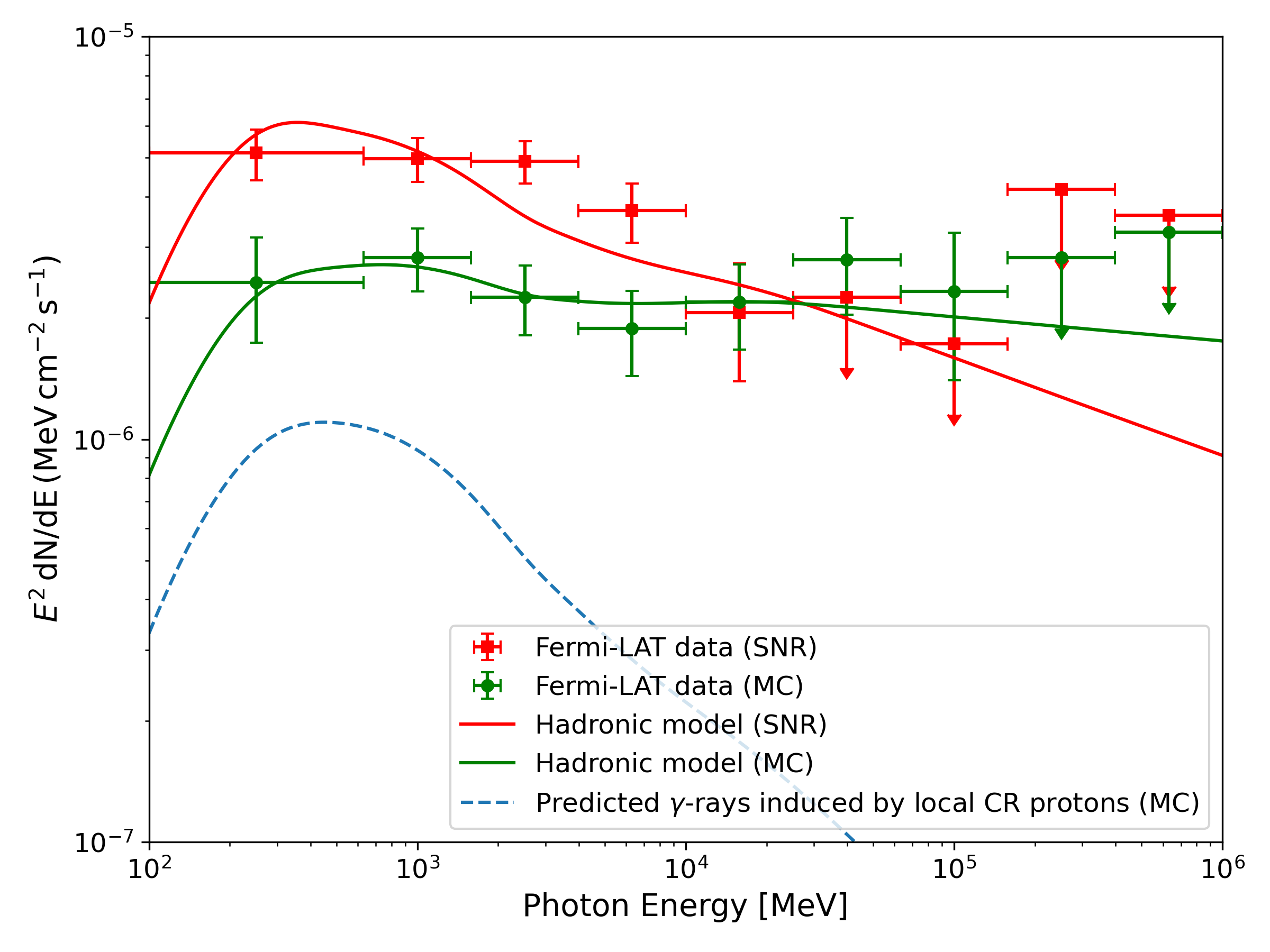}
\caption{SEDs of the $\gamma$-ray emission from S147 region fitted with hadronic models. The red data show the {\sl Fermi}-LAT measurements for the SNR shell (H$\alpha$ component) and the green data represent the emission associated with the MC, i.e. the CO ($0$--$5\,\mathrm{km\,s^{-1}}$) component.
The solid red and green curves show the best-fit hadronic model for $\gamma$-ray spectra associated with the SNR shell and the MC, respectively. For comparison, the blue dashed curve represents the expected $\gamma$-ray emission produced by the local Galactic CR proton \citep{10.1093/mnras/sty2235} interacting with the same MC.
}
\label{fig:pl_fit}
\end{figure}

We performed the numerical integration and spectral fitting using the \texttt{NAIMA} package  \citep{2015ICRC...34..922Z}, which implements the detailed proton-proton interaction cross-sections $\mathrm{d}\sigma/\mathrm{d}E_{\gamma}$ based on \texttt{Pythia8} parameterizations \citep{PhysRevD.90.123014}. The model parameters $J_{0}$ and $\alpha$ were left free and constrained by fitting the observed SEDs, with the target mass $M_{\mathrm{gas}}$ provided as a fixed input. The best-fit parameters for the hadronic model derived from the SED fitting are summarized in Table~\ref{tab:pl_hadronic_fit}. As shown in Figure~\ref{fig:pl_fit}, the hadronic model provides a good description of the experimental data for both the SNR shell and the MC components, reproducing the spectral shape within the uncertainties.

\begin{table}[htbp]
\centering
\caption{Fitting Results of the hadronic model parameters}
\label{tab:pl_hadronic_fit}
\begin{tabular}{lcc}
\hline\hline
Region &
$J_{0}$ &
$\alpha$ 
\\
 &
($\mathrm{GeV^{-1}\,cm^{-2}\,s^{-1}\,sr^{-1}}$) &
 \\
\hline
SNR &
$18.91^{+2.18}_{-1.98}$ &
$2.30^{+0.03}_{-0.02}$ \\
MC &
$0.75^{+0.19}_{-0.16}$ &
$2.12^{+0.06}_{-0.05}$ \\
\hline
\hline
\end{tabular}
\end{table}

\begin{figure}[htbp]
\centering
\includegraphics[width=0.7\textwidth]{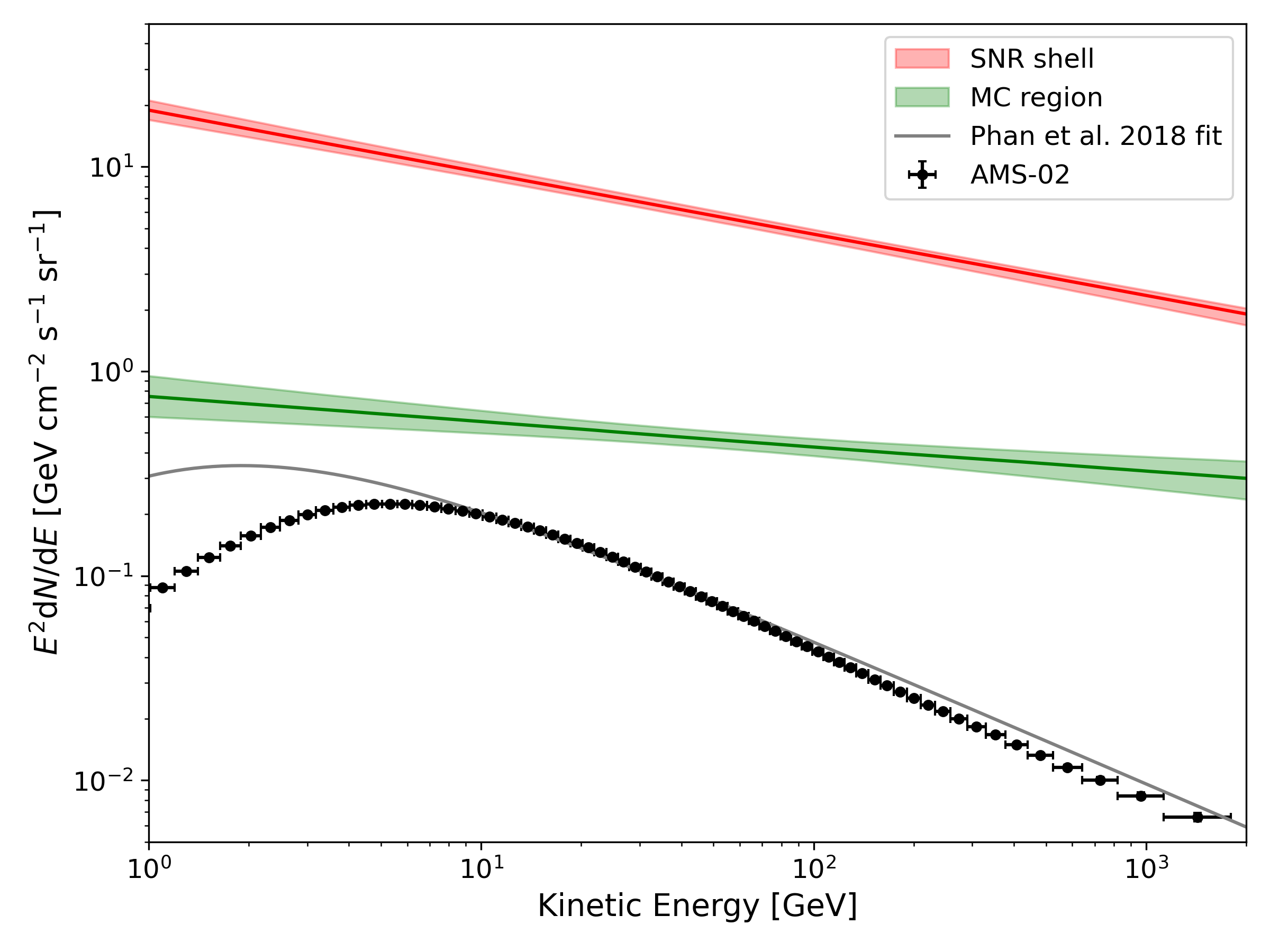}
\caption{Comparison between the inferred CR proton spectra in the vicinity of S147 and local Galactic CR measurements. The red and green curves represent the converted CR proton intensities for the SNR shell and the MC regions, respectively. The shaded bands indicate the 16th–84th percentile ranges derived from the spectral fitting. The black data points with error bars show the locally measured CR proton spectrum reported by AMS-02 \citep{AGUILAR20211}, retrieved from the CRDB database \citep{refId0,universe6080102}. The gray band corresponds to a fit to CR proton data combining Voyager~1 measurements at low energies and AMS observations at high energies, as presented in \citet{10.1093/mnras/sty2235}.}
\label{fig:cr_comparison}
\end{figure}

Beyond reproducing the observed $\gamma$-ray spectra, the hadronic modeling allows us to infer the properties of the CR proton populations responsible for the emission. And we can compare our inferred spectra against the local Galactic CR proton spectrum derived from AMS-02 and Voyager~1 data \citep{AGUILAR20211, 10.1093/mnras/sty2235}. As illustrated in Figure~\ref{fig:cr_comparison}, the derived CR proton intensities in both the SNR shell and the MC regions significantly exceed the local measurements over a broad energy range. We note that this conclusion is robust against parameter uncertainties for both components. For the SNR shell, reconciling the $\gamma$-ray emission with the local Galactic CR background would require a target gas mass orders of magnitude higher than our reference value of $1000~M_{\odot}$, which far exceeds current estimates \citep{Katsuta:2012zz, Khabibullin:2024hbr}. For the MC, even allowing for plausible uncertainties in the distance estimation and the CO-to-H$_2$ conversion factor, it is difficult to account for the observed excess solely by the local CR background. Notably, the inferred spectra are considerably harder ($\alpha \approx 2.1-2.2$) than the local measurements ($\alpha \approx 2.7$). To rigorously test the background contribution, we modeled the expected $\gamma$-ray emission using the locally Galactic CR spectrum as a fixed input. As demonstrated in Figure~\ref{fig:pl_fit} (blue dashed curve), the Galactic CR background alone fails to reproduce the data, significantly under-predicting the high-energy flux.

The detection of a harder and enhanced CR proton population is consistent with the presence of an additional CR component associated with S147. A plausible scenario is that high-energy protons were accelerated by the SNR shock and subsequently escaped into the surrounding MCs. This interpretation is supported by the spectral index of the protons illuminating the clouds ($\alpha \approx 2.12$), which is consistent with the theoretical prediction for the DSA mechanism ($\alpha \approx 2.0$) and significantly flatter than the softened Galactic sea spectrum. Since S147 is a mature SNR, these high-energy particles may have escaped during an earlier evolutionary phase when the shock velocity was higher. These escaping particles have since propagated to the nearby dense gas, producing the observed enhanced $\gamma$-ray emission \citep[e.g.,][]{2026arXiv260422621C}. These results provide observational support for a scenario in which S147 acts as a source of CRs that illuminate the surrounding molecular environment.

\section{Summary and Conclusion}
\label{sec:conclusion}

In this work, we performed a comprehensive study of the $\gamma$-ray emission from the middle-aged SNR S147 using approximately 16.5 years of \textit{Fermi}-LAT data. By utilizing event types with the best angular resolution at low energies and combining the data with CO molecular line surveys, we have resolved the complex morphology of the region. A key result of our analysis is the identification of an extended $\gamma$-ray component that is spatially distinct from the H$\alpha$ filaments of the SNR shell. This emission exhibits a strong correlation with dense MCs identified in the $0$--$5\,\mathrm{km\,s^{-1}}$ velocity interval. Spatial modeling confirms that a CO-based template provides a statistically superior description of this emission compared to geometric models or the unassociated point source previously reported in this region, suggesting that the cataloged point source may reflect a local enhancement within a larger extended structure.

Our spectral analysis indicates that the emission from the SNR shell is well explained by the hadronic interactions of CRs with the ambient gas, consistent with the interpretation in \citet{Katsuta:2012zz}. More importantly, we find that the emission associated with the MCs in the vicinity of S147 exhibits a hard power-law spectrum with an index of $\Gamma \approx 2.1$. The CR intensity derived for this region significantly exceeds the local Galactic background. This enhancement, combined with the hard spectral index, is consistent with a scenario in which the clouds are illuminated by CRs that escaped from S147 during the remnant's earlier, more active evolutionary phases.

Notably, the hard spectrum of the MC component is consistent with extending to hundreds of GeV, showing no significant evidence of a spectral cutoff within current uncertainties. This suggests that the maximum energy of the escaping particles may extend well into the very high energy regime. Although S147 is not included in the first LHAASO catalog \citep{LHAASO:2023rpg}, recent LHAASO observations provide a useful reference for its potential detectability. In particular, the source C1 in the IC~443 region has been detected by LHAASO with a significance of $13.1\sigma$ \citep{LHAASO:2025aoo}. The spectrum of C1 reported in that work shows similarly hard GeV--TeV behavior and a flux level comparable to that of the MC component of S147 in the overlapping energy range. Therefore, if the S147 MC component does not exhibit an early cutoff above the LAT energy range, it could be a potential candidate for detection at TeV energies by LHAASO. Future detection of emission at TeV energies would not only constrain the maximum particle energy achieved during the remnant's early active phase but also establish S147 as a valuable laboratory for studying how middle-aged SNRs continue to impact the surrounding interstellar medium through escaping CRs.

\begin{acknowledgments}
This work is supported by the National Key Research and Development Program of China (2022YFF0503304), the National Natural Science Foundation of China (Nos. 12322302, 12573053), the Project for Young Scientists in Basic Research of Chinese Academy of Sciences (No. YSBR-061), and the Natural Science Foundation for General Program of Jiangsu Province of China (NO.BK20252108, BK20242114).
\end{acknowledgments}

\appendix

\section{Counts map comparison}
\label{app:counts_maps}

\begin{figure}[htbp]
\centering
\includegraphics[width=0.95\textwidth]{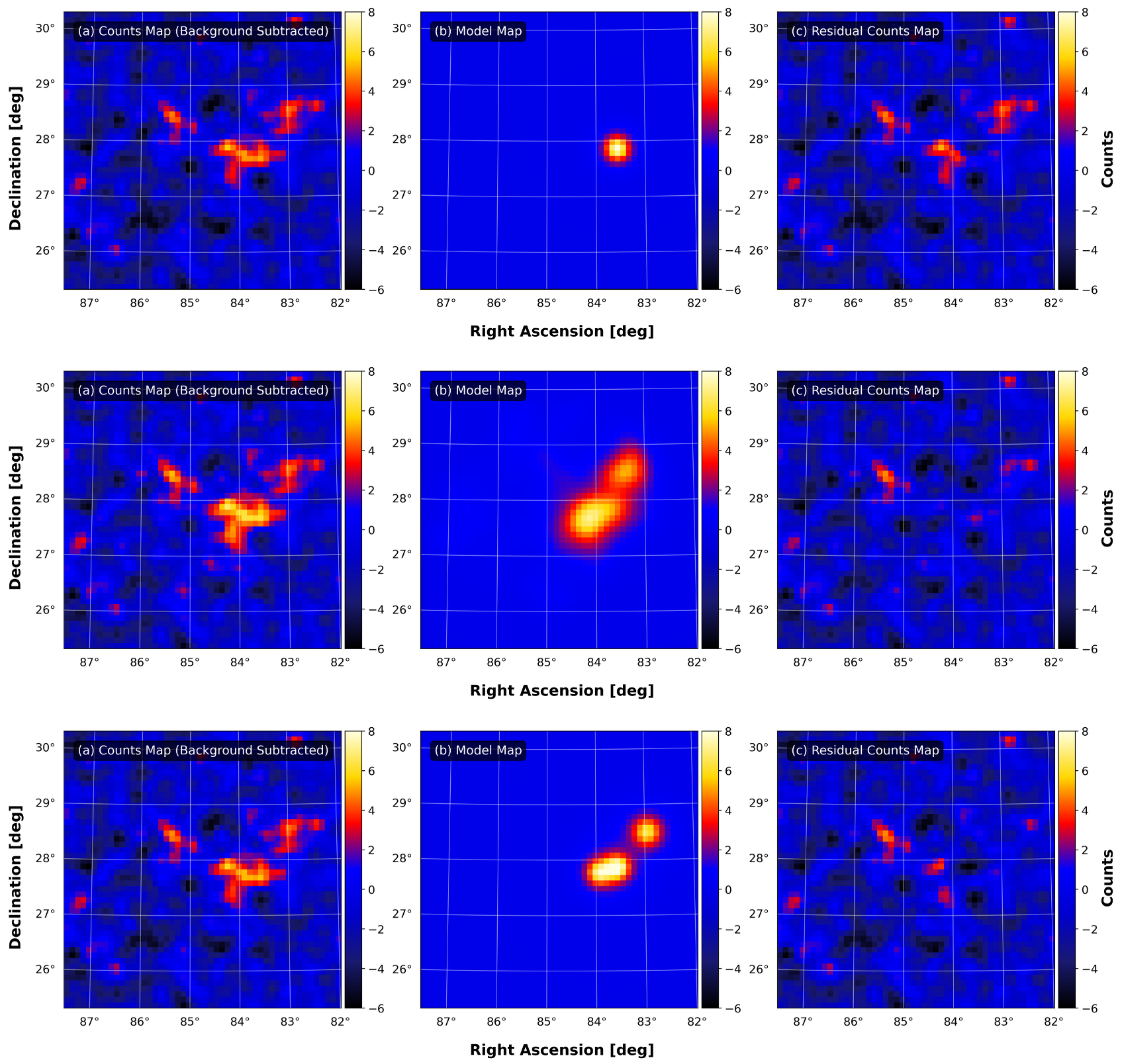}
\caption{Background-subtracted counts maps, model maps, and residual maps (Counts $-$ Model) for the S147 region in the energy range $1~\mathrm{GeV}$--$1~\mathrm{TeV}$ under different spatial models. The three rows show results for the baseline model (top), the best-fit CO model (middle), and the multi-point-source model (bottom), respectively.
}
\label{fig:1gev_counts_maps}
\end{figure}

\begin{figure}[htbp]
\centering
\includegraphics[width=0.95\textwidth]{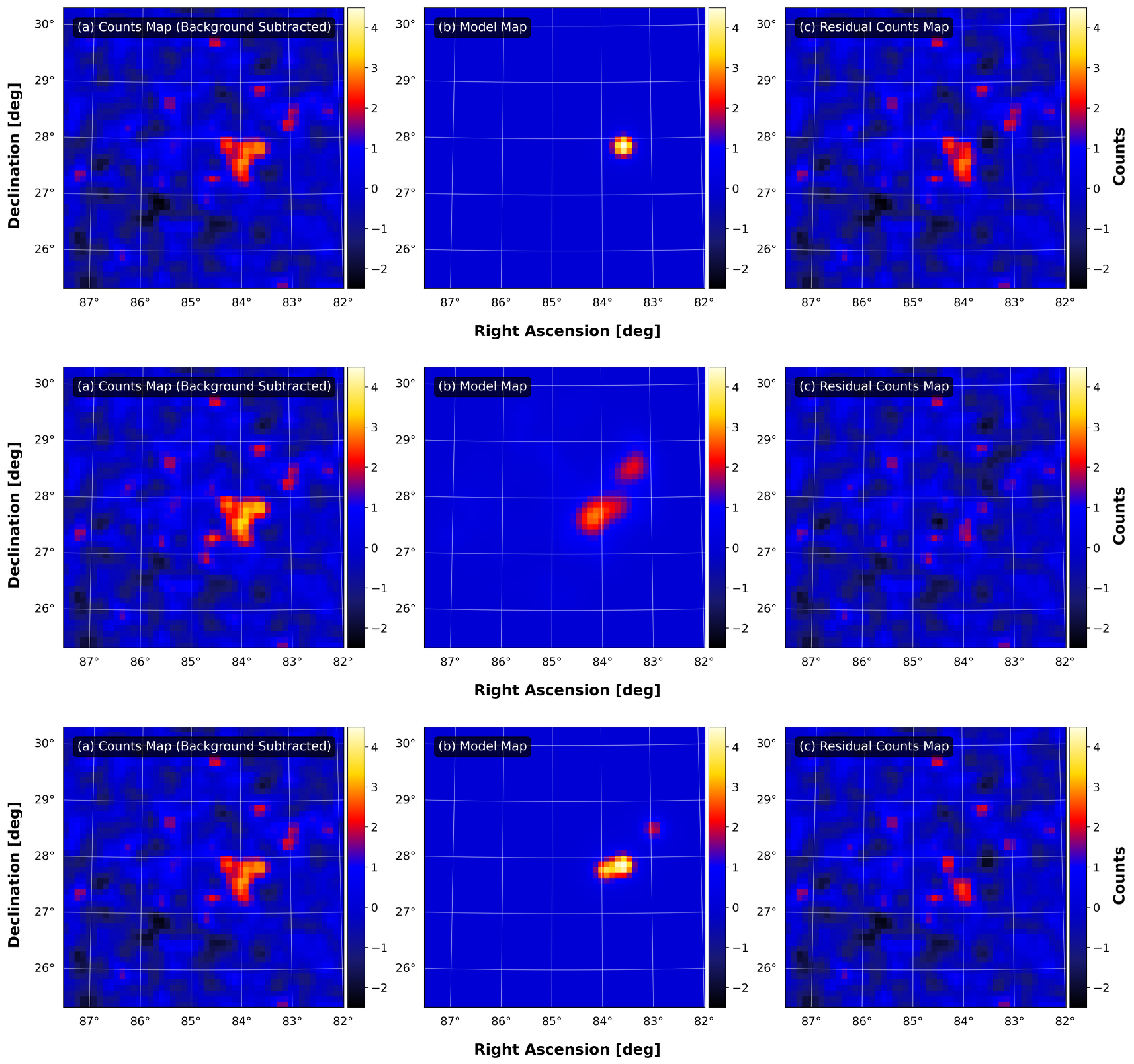}
\caption{Same as Figure~\ref{fig:1gev_counts_maps}, but for the energy range $3~\mathrm{GeV}$--$1~\mathrm{TeV}$.}
\label{fig:3gev_counts_maps}
\end{figure}

As a complementary check, we visually evaluate the performance of different spatial models by comparing the background-subtracted counts maps, model maps, and corresponding residual maps (Counts $-$ Model) for three key competing scenarios: the Baseline model, the best-fit CO model, and the multi-point-source model. The background-subtracted counts maps are obtained by subtracting the fitted background model from the observed counts. In this comparison, the background model is defined as all components of the baseline model except 4FGL~J0534.2+2751. For each tested spatial model, the parameters left free in the likelihood fit are re-optimized before constructing the maps. The multi-point-source model corresponds to the Scenario A case with two additional point sources, which provides the most flexible unresolved-source alternative tested in Table~\ref{tab:model_comparison}. The comparison is performed in the energy range 1~GeV--1~TeV, as presented in Figure~\ref{fig:1gev_counts_maps}. Since the LAT PSF becomes narrower at higher energies, allowing the spatial structure to be better resolved, we further perform the same comparison in the energy range 3~GeV--1~TeV, as shown in Figure~\ref{fig:3gev_counts_maps}. The visual comparison is consistent with the likelihood results. The CO-based template provides a more coherent reproduction of the observed spatial distribution than the Baseline and multi-point-source models, and it leaves smaller residual structures.

\clearpage
\bibliography{cite}{}
\bibliographystyle{aasjournalv7}

\end{document}